\documentstyle[12pt]{article}
	\title{Granular Material Flowing Down an Inclined Chute: A Molecular
Dynamics Simulation
		\vspace*{2ex}}
	\author{Thorsten P\"oschel}
	\date{\small \today}
\begin{document}
\maketitle
\begin{center}
 	preprint HLRZ 22/92 \\
	\vspace*{10ex}
	HLRZ, KFA J\"ulich, Postfach 1913 \\
	D-5170 J\"ulich, Germany  \\
 		\vspace*{1ex}
	and \\
 		\vspace*{1ex}
	Humboldt--Universit\"at zu Berlin \\
	FB Physik, Institut f\"ur Theoretische Physik \\
	Invalidenstra\ss e 42, D-1040 Berlin \\
\end{center}
\vspace*{2ex}
\begin{abstract}
\noindent
	Two-dimensional Molecular Dynamics simulations are used to model the
free surface flow of spheres falling down
 	an inclined chute. The interaction between the particles in our model
is assumed to be subjected to the Hertzian
	contact force and normal as well as shear friction. The stream of
particles shows a characteristic
	height profile, consisting of layers of different types of
fluidization. The numerically observed
	flow properties agree qualitatively with experimental results.
\end{abstract}
\vfill
PACS. numbers: 05.60, 46.10, 02.60

\setcounter{topnumber}{3}
\renewcommand{\topfraction}{1.0}
\newpage
\section{Introduction}
Flows of granular material such as sand or powder reveal macroscopic behaviour
far from the behaviour of flowing
liquids. Examples of astonishing effects are heap formation under vibration
\cite{evesque}-\cite{mehta},
size segregation
\cite{haff}-\cite{rosetto},
formation of convection cells \cite{gallas1}-\cite{taguchi}
and density waves emitted from outlets \cite{bexter} \cite{ristow} or from
flows
through narrow pipes \cite{poschel} and on surfaces
\cite{johnson}-\cite{savage_92}
and displacements inside shear cells \cite{campbell_85}-\cite{jaeger}.
For its exotic behaviour granular materials have been of interest to
physicists over many decades.

Recently highly sophisticated experimental investigations were carried out
concerning the two-dimensional free surface
flow of plastic spheres generated in an inclined glass-walled chute
\cite{drake}. The observed flows at sufficiently low
inclination consists of layers of different types of fluidization. The flow
generally can be divided into frictional and
collisional regions. At least in the case of a rough bed the frictional region
typically consists of a quasi-static zone
at the bottom and above of the so called block-gliding zone where blocks of
coherent moving particles move parallel
to the bed.

In the current paper we want to present the results of our numerical
investigations concerning the problem mentioned
above. Beginning at an initial state of particle positions, velocities and
angular velocities the dynamics of the
system will be determined by integrating Newton's equation
of motion for each particle for all further time steps. In our case the forces
acting upon the particles are exactly known
for each configuration of our system. For this reason we have chosen a sixth
order predictor--corrector
Molecular Dynamics simulation \cite{allen} for the
evaluation of the positions of the particles and a fourth order
predictor--corrector method for the calculation of the
particles angular movement to be the appropriate numerical methods to solve
our problem.
The height profiles of the particle's velocities, of the particle density and
the segregation of the particle flow into
different types of fluidization were the essential points of interest of our
investigations. As shown below the
properties of the particle flow varies sensitively with the smoothness of the
bed surface. As we will demonstrate, the
numerical results are in qualitative accordance with the experimental data.

\section{Numerical Model}

The system consists of $N=400$ equal two-dimensional spherical particles with
radius $R$ moving
down a chute of length $L=50\cdot R$
inclined by an angle ($\alpha = 10^o \ldots 30^o$) which is assumed to have
periodic boundary
conditions in the direction parallel to the bed. Our results did not vary
qualitatively with the
number of particles provided that they are enough to yield reliable statistics
($N\ge 250$).
In simulations with more particles ($N=1000$) we observed the same effects,
theprofiles of the
averaged particle density, the cluster density and the particle velocity
becameslightly
smoother, while the calculation time rose substantially.
In fig.~\ref{setup} the setup of our computer simulations is drawn.
\par
\unitlength1.0cm
\begin{figure}[h]
\begin{picture}(9,6)
	\thicklines
	\put(3,1){\line(1,0){6}}
	\put(3,1){\line(3,1){6}}
	\put(3.2,2){\vector(-3,-1){1}}
	\put(9.2,4){\vector(-3,-1){1}}
	\put(9.3,2.7){\vector(0,-1){1.5}}
	\put(9.4,2.4){\makebox(0,0)[l]{acceleration due to}}
	\put(9.4,2.0){\makebox(0,0)[l]{to gravity}}
	\put(4.0,1.2){\makebox(0,0)[[l]{$\alpha$}}
	\thinlines
	\put(9.0,3.0){\vector(3,1){0.5}}
	\put(9.6,3.2){\makebox(0,0)[l]{x}}
	\put(3,1){\vector(-1,3){0.65}}
	\put(2.2,3.2){\makebox(0,0)[l]{z}}
	\put(9.0,3){\line(-1,3){0.55}}
	\put(5.35,3.3){\vector(-3,-1){2.8}}
	\put(5.7,3.433){\vector(3,1){2.8}}
	\put(5.45,3.4){\makebox(0,0)[l]{L}}
	\put(9.0,3.95){\line(1,2){0.3}}
	\put(9.4,4.8){\makebox(0,0)[l]{periodic boundary}}
	\put(9.4,4.4){\makebox(0,0)[l]{conditions}}
\end{picture}
\caption{\it Computer experimental setup}
\label{setup}
\vspace{2ex}
\end{figure}
The particles were accelerated with respect to the acceleration due to gravity
($ g~=-9.81~m/s^2 $, the mass $M$ was set to unity):

\begin{equation}
	F_Z~=-M\cdot g\cdot \cos(\alpha)
\end{equation}
\begin{equation}
	F_X~=-M\cdot g\cdot \sin(\alpha)
\end{equation}
where $F_Z$ and $F_X$ are the forces acting upon each particle in the
directions normal to the inclined chute and
parallel to it respectively.

The surface of the chute was built up of smaller spherical particles
($R_S=0.66\cdot R$) of the same material as the moving
particles, i.e. with the same elasticity and friction parameters. These
sphereswere arranged at fixed positions
either regularly and close
together in order to simulate a smooth bed or randomly out of the interval
[$2 \cdot R_S; 2 \cdot \sqrt{R \cdot R_S + {R_S}^2}$ ) to
simulate a rough bed (fig.~\ref{bed}).
\par
\unitlength1.0cm
\begin{figure}
\begin{picture}(12,3)
	\thicklines
	\put(3,0.92){\line(1,0){3}}
	\put(7,0.92){\line(1,0){3}}
	\put(3,0.94){\line(1,0){3}}
	\put(7,0.94){\line(1,0){3}}
	\put(3,0.96){\line(1,0){3}}
	\put(7,0.96){\line(1,0){3}}
	\put(3,0.98){\line(1,0){3}}
	\put(7,0.98){\line(1,0){3}}
	\put(3,1){\line(1,0){3}}
	\put(7,1){\line(1,0){3}}
	\put(3,1.02){\line(1,0){3}}
	\put(7,1.02){\line(1,0){3}}
	\put(3,1.04){\line(1,0){3}}
	\put(7,1.04){\line(1,0){3}}
	\put(3,1.06){\line(1,0){3}}
	\put(7,1.06){\line(1,0){3}}

	\put(3.15,1.07){\circle*{0.3}}
	\put(3.45,1.07){\circle*{0.3}}
	\put(3.75,1.07){\circle*{0.3}}
	\put(4.05,1.07){\circle*{0.3}}
	\put(4.35,1.07){\circle*{0.3}}
	\put(4.65,1.07){\circle*{0.3}}
	\put(4.95,1.07){\circle*{0.3}}
	\put(5.25,1.07){\circle*{0.3}}
	\put(5.55,1.07){\circle*{0.3}}
	\put(5.85,1.07){\circle*{0.3}}

	\put(7.15,1.07){\circle*{0.3}}
	\put(7.65,1.07){\circle*{0.3}}
	\put(8.115,1.07){\circle*{0.3}}
	\put(8.42,1.07){\circle*{0.3}}
	\put(8.82,1.07){\circle*{0.3}}
	\put(9.2,1.07){\circle*{0.3}}
	\put(9.7,1.07){\circle*{0.3}}

	\thinlines
	\put(4,1){\line(-1,1){0.5}}
	\put(9,1){\line(1,1){0.5}}
	\put(3.5,1.7){\makebox(0,0)[b]{smooth bed}}
	\put(9.5,1.7){\makebox(0,0)[b]{rough bed}}
\end{picture}
\caption{\it Rough and smooth bed both consists out of spheres with diameter
$R_S=0.66 \cdot R$. In the case of the
smooth bed these spheres are arranged close together while in the case of the
rough bed they are arranged irregularely.}
\label{bed}
\vspace{2ex}
\end{figure}
To describe the interaction between the particles and between the particles
and the bed we applied the ansatz
described in~\cite{cundall} and slightly modificated in~\cite{haff}:
\begin{equation}
	\vec{F}_{ij} =  \left\{ \begin{array}{cl}
		F_N \cdot \frac{\vec{x}_i - \vec{x}_j}{|\vec{x_i}-\vec{x}_j|}
+F_S \cdot  \left({0 \atop 1} ~{-1 \atop 0} \right) \cdot
                  \frac{\vec{x}_i - \vec{x}_j}{|\vec{x_i}-\vec{x}_j|}
		& \mbox{if $|\vec{x_i}-\vec{x}_j| < 2 \cdot R$} \\[1ex]
		0
& \mbox{otherwise}
		\end{array}
		\right.
\end{equation}

with

\begin{equation}
	F_N  = k_N \cdot (r_i + r_j - |\vec{x}_i -
\vec{x}_j|)^{1.5}~+~\gamma_N\cdot M_{eff}\cdot (\dot{\vec{x}}_i -
\dot{\vec{x}}_j)
\end{equation}

and

\begin{equation}
	F_S = \min \{- \gamma_S \cdot M_{eff} \cdot v_{rel}~,~ \mu \cdot |F_N|
\}
\label{eq_coulomb}
\end{equation}

where
\begin{equation}
	v_{rel} = (\dot{\vec{x}}_i - \dot{\vec{x}}_j) + R \cdot
(\dot{\Omega}_i - \dot{\Omega}_j)
\end{equation}
\begin{equation}
	M_{eff} = \frac{M_i \cdot M_j}{M_i + M_j} = \frac{M}{2}
\end{equation}
\vspace{2ex}
\par
\noindent
The terms $\vec{x}_i$, $\dot{\vec{x}}_i$, $\dot{\Omega}_i$ and $M_i$ denote
the current position, velocity,
angular velocity and mass of the $i$th particle. The model includes an elastic
restoration force
which corresponds to the microscopic assumption that the particles can
penetrate slightly each other. In order
to mimic three--dimensional behaviour this
force rises with the power $\frac {3}{2}$ which is called {\sc ``Hertz}ian
contact force'' \cite{landau}. The other terms
describe the energy dissipation of the system due to collision between
particles according to normal and
shear friction. The parameters $\gamma_N$ and $\gamma_S$ stand for the normal
and shear friction
coefficients. Eq.~\ref{eq_coulomb} takes into account that the particles will
not transfer
rotational energy but slide upon each other in
case the relative velocity at the contact point exceeds a certain value
depending on the
normal component of the force acting between the particles ({\sc
Coulomb-}relation \cite{coulomb}.   For all our
simulations we have
chosen the constant parameters $\gamma_N=1000 s^{-1}$, $\gamma_S=300 s^{-1}$,
$k_N=100 \frac{\mbox{N}}{\mbox{m}^{1.5}}$ and
 $\mu=0.5$.
For the integration time step we used
$\Delta t = 0.001 s$ to provide numerical accuracy. When doubling the time
step we got exactly the same results.

\section{Results}
\subsection{Initialization}
In the simulations the particles are placed initially on a randomly disturbed
lattice above the bed
so that they do not touch each
other in the beginning.
In order to obtain periodic boundary conditions of the bed we have to fit the
length of the bed in a way that the center
of the first sphere of the bed is at $x=0$ and can reappear at $x=L$ without
colliding with its neighbours.
Suppose the length of the chute was chosen to be $L_0$ it was fitted to
$L$ with $L \in [L_0, L_0+R_S)$ to hold periodic boundary conditions. The
particles' initial velocities
were zero.  After starting the simulation the particles gain
kinetic energy since they fall down freely without any collisions, i.e.
without dissipation. This is the reason for
a peak of the kinetic energy
in the very beginning of the simulation.

\subsection{Smooth Bed}
For the case of the smooth bed fig.~\ref{E_smooth} shows the kinetic energy of
the system. The chute was
inclined by an angle $\alpha=20^o$. The curve shows a significiant kink which
separates two different regimes of the
particle movement as shown below. The numerical simulations reported in 3.2.
and 3.3. run while the particles are
accelerated due to gravity, their kinetic energy rises. After waiting enough
time the system comes into saturation
and finds its steady state as we will demonstrate in 3.4. below.
In fig.~\ref{snap_smooth_slow} a snapshot of the grains' movement before the
kink of the
energy is displayed.
The small line drawn within each particles shows the direction and the amount
of the velocity of the particle,
its length was scaled due to the fastest particle. In all our simulations the
rotational energy is negligible.
In fig.~\ref{E_smooth} the
evolution of the rotational energy coincides with the time-axis.
\par
\begin{figure}
\vspace{2ex}
\caption{\it Evolution of the kinetic (transversal) and rotational energy for
the particle flow on a smooth bed. The
sharp kink separates two different regimes of the particle movement called the
low energy regime and the high energy regime.}
\label{E_smooth}
\vspace{2ex}
\end{figure}
\begin{figure}
\vspace{2ex}
\caption{\it Snapshot after 90.000 timesteps within the low energy regime.
Particles which belong to a cluster are drawn bold
faced. All grains at the bottom ($h<100$) form one large block.
Only few off them above the block belong to clusters.}
\label{snap_smooth_slow}
\vspace{2ex}
\end{figure}
As indexed by the velocity arrow
drawn inside the particles the flux divides into the lower block gliding zone,
where nearly all of the particles move coherently
with equal velocity forming a single block of particles. The particles which
belong to one block are drawn bold faced.
More precisely to
identify which particles form a block we applied the following rule: Each two
particles moving in the neighbourhood
($|\vec{r_i}-\vec{r_j}|<\lambda \cdot R$) for a time of at least 5000 time
steps which corresponds to a covered distance of
$(10\ldots 30) \cdot R$
approximately belong to the same block. If $\lambda$ was chosen very small
($\lambda <2.1$) then
even very small elastical bouncing of two neighboured particles leads to the
wrong interpretation that the
particles are not neighboured anymore while two particles moving with a small
relative velocity
were detected to belong to a cluster when $\lambda$ was chosen too large. With
the value $\lambda =2.2$ we got
correct results.
As visible the grains in the lower part move coherently as a block in the case
of
low energy distinct from the high energy behaviour as shown below in
fig.~\ref{snap_smooth_fast}.
The upper particles move irregularely.
\par
Figs. \ref{smooth_slow_dichte}, \ref{smooth_slow_vx} and \ref{smooth_slow_cl}
show the particle-density, velocity in
horizontal direction due to the bed and cluster-density
depending on the vertical distance from the bed $h$ corresponding to the
snapshot given in fig.~\ref{snap_smooth_slow}.
\par
\begin{figure}
\vspace{2ex}
\caption{\it Particle density depending on the vertical distance from the bed
$h$
(low energy regime). Within the block
the partical density does not vary.}
\label{smooth_slow_dichte}
\vspace{2ex}
\end{figure}
\begin{figure}
\vspace{2ex}
\caption{\it Averaged horizontal velocity (low energy regime). The grains
which belong to the block move with equal
velocity, i.e. there is really only one block but not different blocks gliding
on each other. The velocity changes
desultory at the upper boundary of the block ($h\approx 100$), the free moving
particles are 1.3 to 1.5 times as fast as the block.}
\label{smooth_slow_vx}
\vspace{2ex}
\end{figure}
\begin{figure}
\vspace{2ex}
\caption{\it Cluster density (low energy regime). Above the block the cluster
density vanishes. There are almost no
particles which belong to a cluster except of the block at the bed.}
\label{smooth_slow_cl}
\vspace{2ex}
\end{figure}
The velocity distribution in horizontal direction separates into two regions,
the lower particles including the block and the
grains close to the block move with nearly the same velocity. The particles on
the surface of the flow move with larger velocity. There is a sharp transition
between the velocities of particles which
belong to the block and the grains moving at the surface. The term cluster
density means the portion of particles which
belong to one of the clusters.
The cluster density as a function of the distance from the bed $h$ does not
vary from the bed to the top of the block and
lowers then fastly as expected from the snapshot (fig.\ref{snap_smooth_slow})
because there are almost no smaller
clusters except the block. The velocities in fig.~\ref{smooth_slow_vx}
have negative sign
because the particles
move from the right to the left i.e. in negative direction.
\par
As pointed out above the  sharp kink of the energy in fig.~\ref{E_smooth}
corresponds to different regimes.
Figs.~\ref{snap_smooth_fast}-\ref{smooth_fast_cl} are
analogous to figs.~\ref{snap_smooth_slow}-\ref{smooth_slow_cl} for a snapshot
short time after the transition into the
high velocity regime at time step 110.000.
\par
\begin{figure}
\vspace{2ex}
\caption{\it Snapshot after 110.000 timesteps in the high energy regime. The
block
(fig.~4)
separated within a
relatively short time into many independently moving smaller clusters.
The rest of the block lies still close to the bed.
It also vanishes after some time.}
\label{snap_smooth_fast}
\vspace{2ex}
\end{figure}
\begin{figure}
\vspace{2ex}
\caption{\it Particle density depending on the vertical distance from the bed
$h$ (high energy regime). Although
the properties of the flow are very different from the flow in the low energy
regime the profile of the particle density
differs not significiantly.}
\label{smooth_fast_dichte}
\vspace{2ex}
\end{figure}
\begin{figure}
\vspace{2ex}
\caption{\it Averaged horizontal velocity (high energy regime). It is the most
noticeable index to the fact that the particle
flow changed into an other regime.}
\label{smooth_fast_vx}
\vspace{2ex}
\end{figure}
\begin{figure}
\vspace{2ex}
\caption{\it Cluster density (high energy regime). Between the rest of the
block gliding on the chute and the free moving
particles on the top of the flow lies a region of constant cluster-density.}
\label{smooth_fast_cl}
\vspace{2ex}
\end{figure}
If the energy rises over a critical value while the grains move accelerated
due to gravity the particle flow does not
longer divide into the block and the free moving particles but the block
separates  into independently moving
smaller clusters.
This transition occures within a short energy interval which corresponds to a
short time from the beginning of the
simulation.
As shown in
fig.~\ref{smooth_fast_dichte} the density as a function of the distance from
the bed $h$ varies only slightly from
the same function at low energy
(fig.~\ref{smooth_slow_vx}). Nevertheless the profile of the horizontal
velocity
(fig.~\ref{smooth_fast_vx})
differs significiantly from the corresponding profile in the low energy regime
(fig.~\ref{smooth_slow_vx}).
The velocity of the grains rises continually from the bed to the top of the
partical flow. The profile of the
horizontal velocity shows convincing the transition into another regime of
movement. Between the small block of
grains near the chute and the freely moving particles in the collisional zone
there is a region with approximately
constant cluster density (fig.\ref{smooth_fast_cl}). This behaviour is also
observed in the experiment \cite{drake}.
Neighboured clusters move with relative velocities between each other.

\subsection{Rough Bed}
In this section we want to discuss the results of the simulations with the
same set of parameters as in the paragraph above
but with the rough bed. As shown in fig.~\ref{rough_E} the energy-time-diagram
for this case does not show the same behaviour as
for the case of
the rough bed but has roughly a parabolic shape. This shape is maintained
qualitatively over a long time
corresponding to a wide range of kinetic energy of the particles until the
kinetic energy comes into the region of saturation.
The problem of energy saturation will be discussed more detailed below in
section 3.4.
\par
\begin{figure}
\vspace{2ex}
\caption{\it Evolution of the kinetic (transversal) and rotational energy for
the case of the rough bed. The energy
shows no kink as in the flow on the smooth bed (fig.~4) but its shape is close
to parabolic over a wide range.}
\label{rough_E}
\vspace{2ex}
\end{figure}
Corresponding to the last section fig.~\ref{snap_rough} shows a snapshot of a
typical situation of the flow,
figs.~\ref{rough_dichte}-\ref{rough_cl} show the
particle-density, the averaged velocity in horizontal direction and the
cluster-density as a function of
 the distance from the bed $h$.
\vspace{2ex}
\par
\begin{figure}
\vspace{2ex}
\caption{\it Snapshot after 80.000 timesteps. As in fig.~4 the particles which
belong to one of the blocks are drawn bold.}
\label{snap_rough}
\vspace{2ex}
\end{figure}
\begin{figure}
\vspace{2ex}
\caption{\it Particle density depending on the vertical distance from the bed
$h$. It shows no very different behaviour
than the corresponding figures (figs.~5,9) for the case of the smooth bed.}
\label{rough_dichte}
\vspace{2ex}
\end{figure}
\begin{figure}
\vspace{2ex}
\caption{\it Averaged horizontal velocity. The flow consists of different
regions: the slow particles close to
the chute, the block gliding zone and the collisional zone near the top of the
flow. The velocity rises continually
with the distance from the bed $h$.}
\label{rough_vx}
\vspace{2ex}
\end{figure}
\begin{figure}
\vspace{2ex}
\caption{\it Cluster density. There is a zone of significiant higher cluster
density. This region will be called block
gliding zone.}
\label{rough_cl}
\vspace{2ex}
\end{figure}
The snapshot fig.~\ref{snap_rough} shows the situation as it can be observed
over a wide range of energy values.
These layers were
found experimentally \cite{drake}. The flow divides into three different zones
(fig.~\ref{rough_vx}: the slow particles
close to the chute
which velocity rises continously with the distance from the chute $h$, the
particles in the block gliding zone which have
approximately the same velocity and few fast moving particles in the
collisional zone at the top of the flow. Also
fig.~\ref{rough_cl} shows evidently that the flow divides into three regions
of different types of fluidization. The
close to the bed slow particles and the fast particles in the collisional zone
are separated by a zone, where
clusters occure. This zone we call block gliding zone. It was first
investigated experimentally in \cite{drake}.

\subsection{Steady State}
The numerical experiments shown above were accomplished in the accelerated
regime. Since our model includes
friction the system will of course reach a steady state after waiting long
enough time. For the parameter sets
we used in the simulations above this steady state is incident to very high
particle velocities. To obtain
numerical accurate results in the steady state regime therefore one has to
choose a very high time resolution i.e.
a very small time step $\Delta t$. Hence the computation time rises strongly
with the kinetic energy of the system.
\par
To demonstrate that our model is able to reach the steady state regime we have
chosen
the angle of inclination $\alpha = 10^o$ and the time step $\Delta t=0.001$.
Fig.~\ref{steady_satur} shows
the evolution of the energy of the system as a function of time.
\vspace{2ex}
\par
\begin{figure}
\vspace{2ex}
\caption{\it Saturation of the kinetic energy after long time for a flow of
particles down a chute which was inclined by
an angle of $\alpha = 10^o$ only.}
\label{steady_satur}
\vspace{2ex}
\end{figure}
\section{Discussion}
The flow of granular material on the surface of a inclined chute was
investigated through a two dimensional Molecular
Dynamics
simulation. For the case of a smooth bed simulated by regularly arranged small
spheres on the surface of the chute we
found two qualitatively different regimes of particle movement corresponding
to low and high kinetic energies.
The velocity as well as the density profiles agree with experimental results
\cite{drake}. For the case of the
rough bed we did not find such a behaviour. As in the case of the smooth bed
the density and velocity profiles agree with
the measured data. In both cases we observed coherent motion of groups of
grains, so called clusters which also has been
observed in the experiment. Of particular interest seems us to be that the
behaviour of the flow on a chute with
regular (smooth) surface differs significiantly from the flow on a chute with
irregular (rough) surface.
\par
By simulating a system with a smaller inclination angle we gave evidence that
our numerical model is able to reach a steady state at a constant kinetic
energy.
\par
The numerical results show that our two dimensional ansatz for the force
acting on the grains which does not include static
friction is able to describe the experimentally measured scenario
qualitatively correctly for the system considered.
\par
\vspace{5ex}
\begin{flushleft}
	{\large\bf Acknowledgements}
\end{flushleft}
	The author wants to express his gratitude to Hans Herrmann for the
invitation to the HLRZ of the KFA J\"ulich
and to him as well as to Stefan Soko{\l}owski for many fruitful and patient
discussions. I thank Gerald Ristow for his
support concerning the technical problems of Molecular Dynamics.
\vspace{5ex}
\input{cyracc.def}
\newfont{\cyrfnt}{mcyr10}
\newcommand{\cyr}{\baselineskip12.5pt \cyrfnt \cyracc}

\newpage
\listoffigures
\end{document}